          [2001/04/25 1.1 (PWD)]
\documentclass[a4paper,twocolumn]{esapub} 
\usepackage{natbib,paralist,psfrag,color,graphicx,epsfig}
\usepackage[latin1]{inputenc}

\title{Revealing the nature of the highly obscured galactic source IGR J16318$-$4848}
\author{S. Chaty}
\affil{Universit\'e Paris 7 Denis-Diderot, 2 place Jussieu,
F-75251 Paris Cedex 05, \& Service d'Astrophysique, France}
\author{P. Filliatre}
\affil{Service d'Astrophysique \& Fédération APC, CEA/DSM/DAPNIA/SAp, 
F-91191 Gif-sur-Yvette cedex France}

 \def\micron{\mbox{ } \mu \mbox{m}}
\def\arcdeg{^{\circ}}
\def\arcmin{^\prime}
\def\arcsec{''}
\def\hsecp{{\rlap.}^{s}}

\begin{document}


\maketitle

\begin{abstract}
The X-ray source IGR~J16318$-$4848 was the first source discovered by INTEGRAL on 2003, January 29. The high energy spectrum exhibits such a high column density that the source is undetectable in X-rays below 2 keV. On 2003, February, 23--25 we triggered a Target of Opportunity (ToO) program using the EMMI and SOFI instruments on the New Technology Telescope of the European Southern Observatory (La Silla) to get optical and near-infrared (NIR) observations. We discovered the optical counterpart, and confirmed the already proposed candidate in the NIR. We report here photometric measurements in the $R$, $I$ and $J$ bands, upper flux limits in the $Bb$ and $V$ bands, lower flux limits in the $H$ and $K_{s}$ bands. We also got NIR spectroscopy between $0.95$ and $2.52\,\micron$, revealing a large amount of emission lines, including forbidden iron lines and P-Cygni profiles, and showing a strong similarity with CI Cam, another strongly absorbed source. Together with the Spectral Energy Distribution (SED), these data point to a high luminosity, high temperature source, with an intrinsic absorption greater than the interstellar absorption, but two orders of magnitude below the X-ray absorption. We propose the following picture to match the data: the source is a High Mass X-ray binary (HMXB) at a distance between 0.9 and 6.2 kpc, the optical/NIR counterpart corresponds to the mass donor, which is an early-type star, probably a sgB[e] star, surrounded by a dense and absorbing circumstellar material. This would make the second HMXB with a sgB[e] star as the mass donor after CI Cam. Such sources may represent a different evolutionary state of X-ray binaries previously undetected with the lower energy space telescopes; if it is so, a new class of strongly absorbed X-ray binaries is being unveiled by INTEGRAL.
\end{abstract}

\thanks{\it \small Based on observations collected at the European Southern Observatory, Chile (proposal ESO N$°$ 70.D-0340).}

\section{Introduction}

IGR~J16318$-$4848 was the first new source discovered by the INTEGRAL
IBIS/ISGRI imager \citep{ubertini,lebrun}. The source was detected on
2003 January 29 in the energy band 15--40 keV with a mean 20--50 keV
flux of $6\times 10^{11}\,\rm erg\,cm^{-2}\,s^{-1}$, $0.5\arcdeg$
south from the galactic equator \citep{courvoisier}.  The source was
thereafter regularly observed for two months. No X-ray counterpart
could be found in the ROSAT All Sky Survey \citep{voges}. The source
was observed by XMM-Newton on 2003 February 10, which detected a
single X-ray source within the INTEGRAL error box using the EPIC PN
and MOS cameras \citep{struder,turner}, giving the most precise
position to date: $\alpha=16^{\rm h} 31^{\rm m} 48\hsecp6$,
$\delta=-48\arcdeg 49\arcmin 00\arcsec$ with a $4\arcsec$ error box
\citep{schartel}. X-ray spectroscopy revealed a very high absorption
column density: $N_{\rm H}> 10^{24}\,\rm cm^{-2}$
\citep{matt,walter03,walter04}, which renders the source invisible
below 2 keV. This amount of absorption is unusual in Galactic
sources. This could explain the non detection by ROSAT, although the
source was discovered at a similar flux level in archival ASCA
observations in 1994 \citep{murakami} on both GIS and SIS instruments
(between 0.4 and 10 keV). Relatively bright and highly absorbed
sources like IGR~J16318$-$4848 could have escaped detection in past
X-ray surveys and could still contribute significantly to the Galactic
hard X-ray background in the 10--200 keV band.  Based on the broad
high-energy spectral and variability characteristics, Walter et
al. (2004) suggest that IGR~J16318$-$4848 is an X-ray binary, the low
temperature and presence of cutoff at low energy suggesting that the
compact object is a neutron star.\\ The high column density prompted
counterpart research in near-infrared (NIR): within the EPIC error
box, a possible counterpart was proposed by Foschini et al. (2003)
using the Two Micron All Sky Survey (2MASS) with the following
magnitudes: $J=10.2$, $H=8.6$, $K_{s}=7.6$ with an uncertainty of
$\pm0.3$ mag \citep{walter03}. On the other hand, no radio emission at
the position of the source could be detected.  In the course of a
Target of Opportunity (ToO) program at the European Southern
Observatory (ESO) dedicated to look for counterparts of high energy
sources newly discovered by satellites including INTEGRAL (PI
S. Chaty), we carried out on 2003, February, 23-25 photometric
observations in the optical and NIR, and spectroscopic observations in
the NIR with EMMI and SOFI instruments on ESO/NTT. The goals were to
search for likely counterparts within the EPIC error box, to get
informations about the environment and the nature of the source,
especially about the mass donor.\\ In the following, we describe
briefly our observations, report our main results on astrometry,
photometry, spectral energy distribution, absorption and temperature
estimates of the most likely candidate.  We also summarize our results
on spectral lines, distance to the source, and conclude on its nature
and the nature of its components.  The reader should consult Filliatre
and Chaty (2004) for more details (and Chaty and Filliatre 2004 for a
condensed version).

\section{Observations and Results}
\label{secconc}

The source IGR~J16318$-$4848 was the first source discovered by INTEGRAL. In the course of a ToO program using the NTT telescope, we performed  photometric and spectroscopic observations less than one month after its discovery in the optical and NIR domains. 
We list here the main results:

\begin{itemize}

\item We discovered the optical counterpart and confirmed  an already proposed NIR candidate (see Walter et al., 2003), by performing an independent astrometry
for this candidate, giving the following position for the candidate:
$\alpha = 16^{\rm h}31^{\rm m}48\hsecp3$ 
$\delta = -48\arcdeg49\arcmin01\arcsec$.
The optical/NIR images and spectra are shown in Figures \ref{fig:f4} and
\ref{fig:f6}.

\item We obtained photometric measurements for the $R$, $I$ and $J$ bands, and got flux upper limits for $B$ and $V$, flux lower limits for $H$ and $K_{s}$:
$Bb$	$>25.4\pm 1$,          $V$ $>21.1\pm 0.1$,
$R=$	$17.72\pm 0.12$,       $I=$ $16.05\pm 0.54$
$J=$	$10.33\pm 0.14$,       $H$ $<10.35\pm 0.15$
$K_{s}$	$<9.13\pm 0.10$.

\item We derived the absorption towards the source along the line of sight: 
Av$\sim 17.4$ magnitudes, greater than interstellar 
(11.8 mag, see Fig. \ref{fig:f11}) but 2 orders
of magnitude lower than in X-rays.

\item With the continua of our GBF (Grism Blue Filter) and 
 GRF (Grism Red Filter) spectra, our photometric measurements and with
 X-ray, radio and archival data, we constructed a SED, shown in
 Fig. \ref{fig:f12}, covering 10 decades in wavelength.

\item From this SED, we derive 
that the companion star must be a high luminosity, high
temperature star: above $10\,000\,\rm K$.

\item No radio emission associated with a low/hard X-ray state suggests
that the compact object is a neutron star.

\item The distance of the source is constrained between $0.9$ and $6.2$ kpc.

\item The $0.95-2.52\,\micron$ NIR spectra, shown in Figures
\ref{fig:f15}, \ref{fig:f16} and \ref{fig:f17}, are highly unusual, 
very rich in emission lines
(we identified 72 emission lines), including forbidden iron lines and P-Cygni profiles. The lines in the spectrum show no cosmological redshift,
confirming that the object is galactic.

\item 80~\% of these lines have been detected in CI Cam, suggesting a similar nature.

\item These spectra favor the existence
of a highly complex, stratified and dense circumstellar environment, 
with stellar wind or enveloppe.

\item The type of the companion star is consistent with
the position on the colour-magnitude HR diagram, computed for
various absorption values and distances (see figure \ref{fig:f21}).

\item Study of the spectral lines, of the SED and of the CMD
suggest a sgB[e] star so the system
would be a HMXB, probably hosting a neutron star,
like CI Cam. 
This HMXB hosting a sgB[e] star is the most likely hypothesis;
 it would then be the second case after CI Cam.

\end{itemize}

Complementary observations are needed in order to confirm our results, among them we propose:
\begin{itemize}
\item high resolution NIR spectroscopy, if possible extended to optical, in order to:
\begin{itemize}
\item extend the SED in the optical to directly see a NIR excess;
\item check if the similarity with CI Cam observed in the NIR is still valid in the optical; however this will be difficult because of the absorption;
\item improve our results concerning P-Cygni profiles, and line broadening;
\end{itemize}
\item long term follow-up spectroscopy and photometry, in order to:
\begin{itemize}
\item seek for line variability;
\item seek for a periodic behaviour to infer the orbital elements.
\end{itemize}
\item simultaneous multi-wavelength observations, in order to
better understand the nature and geometry of the system
\end{itemize}
This source shows many unusual features, the first being its strong intrinsic absorption. Interestingly, among the ten sources that INTEGRAL has discovered in this region, this feature is common (at least in the X-rays), to the three sources discussed by Revnivtsev 2003: IGR~J16318$-$4848, IGR J16320-4751 and IGR J16358-4726, although the $N_{\rm H}$ column density is lower by an order of magnitude in the two latter systems \citep{rodriguez,patel}. However, a clear identification for the optical/NIR counterpart has been done only for IGR~J16318$-$4848. Moreover, the type of the mass donor, as inferred from our study, has been considered up to now as very rare. There is therefore the possibility that INTEGRAL, with the discovery of IGR~J16318$-$4848, has unveiled a new class of 
obscured high energy binaries (see also Walter et al. 2004). 
This class will deserve much attention in the future,
because they might help us to understand 
the evolution of high-energy binary systems.

\begin{small}
\begin{figure}
\centering
  \includegraphics[width=\columnwidth]{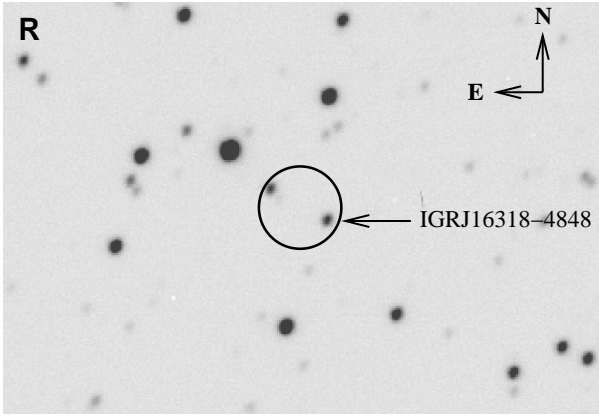}
  \caption{EMMI R band image of the field of view of IGR J16318$-$4848.
We reported the XMM error box of $4\arcsec$.
North is up, east is left. The scale is given by the error box. 
The most likely candidate is at the south-west border of the circle as
indicated by the arrow.
}
  \label{fig:f4}
\end{figure}
\end{small}

\begin{small}
\begin{figure}
  \includegraphics[width=\columnwidth]{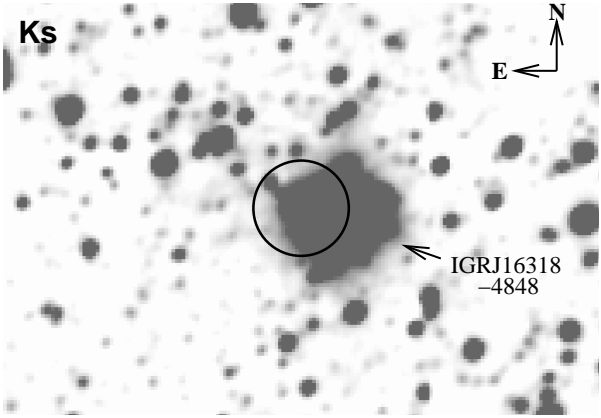}
  \caption{Ks band image of the same field.}
  \label{fig:f6}
\end{figure}
\end{small}

\begin{small}
\begin{figure}
  \includegraphics[width=\columnwidth]{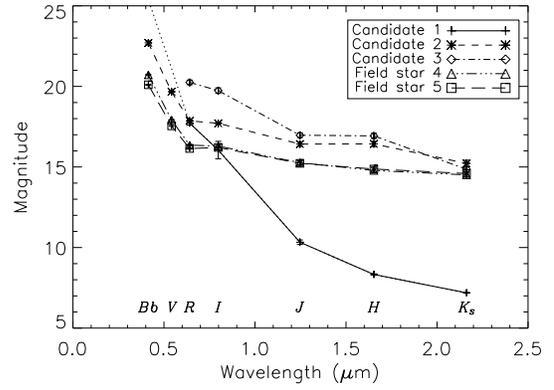}
  \caption{Magnitudes of IGR J16318-4848 (called here candidate 1) 
and field stars}
  \label{fig:f11}
\end{figure}
\end{small}

\begin{small}
\begin{figure}
  \includegraphics[width=6.5cm,angle=90]{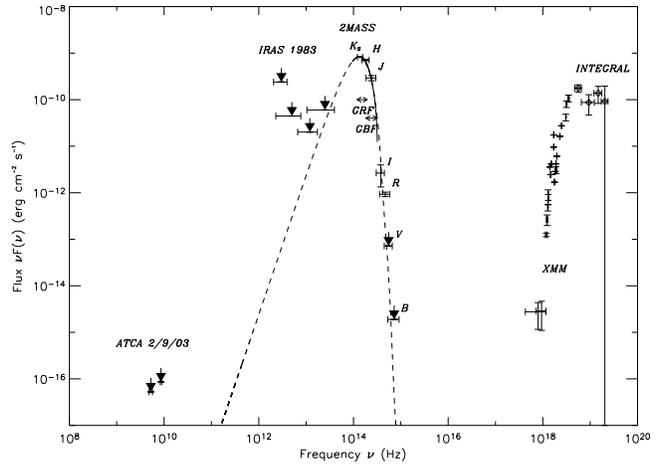}
  \caption{Observed SED of IGR J16318-4848 in $\left(\nu,\,\nu F(\nu)\right)$ units, including the results of our photometry, our rescaled continuum 
GBF and GRF spectra, and literature data. The $B$ and $V$ data are upper limits only. The dashed curve corresponds to an absorbed black body, representing well the data. The results of INTEGRAL, XMM, IRAS and ATCA are also shown.
}
  \label{fig:f12}
\end{figure}
\end{small}

\begin{small}
\begin{figure}
  \includegraphics[width=\columnwidth]{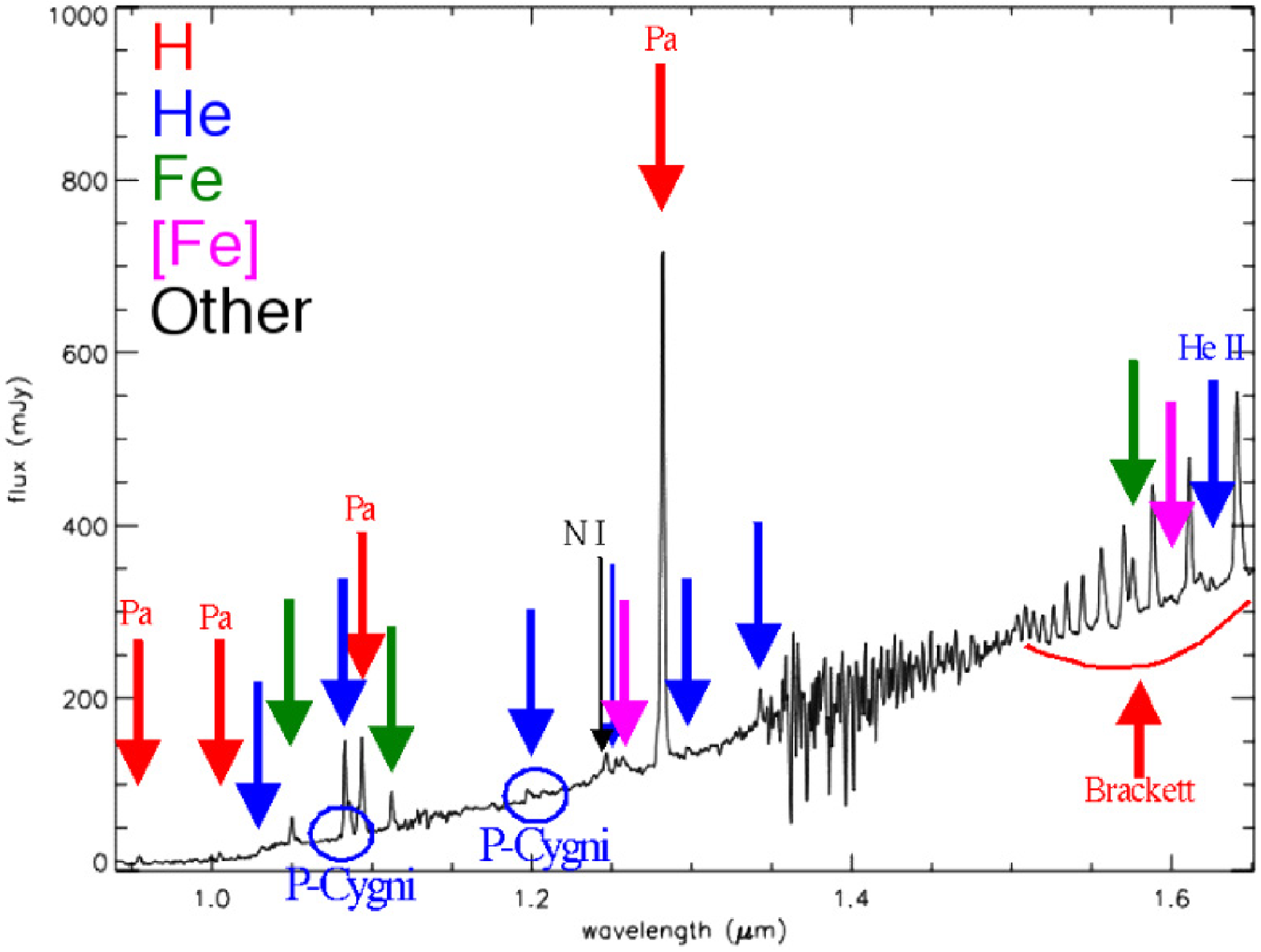}
  \caption{NIR GBF spectrum (0.95-1.65 $\micron$)}
  \label{fig:f15}
\end{figure}
\end{small}

\begin{small}
\begin{figure}
  \includegraphics[width=\columnwidth]{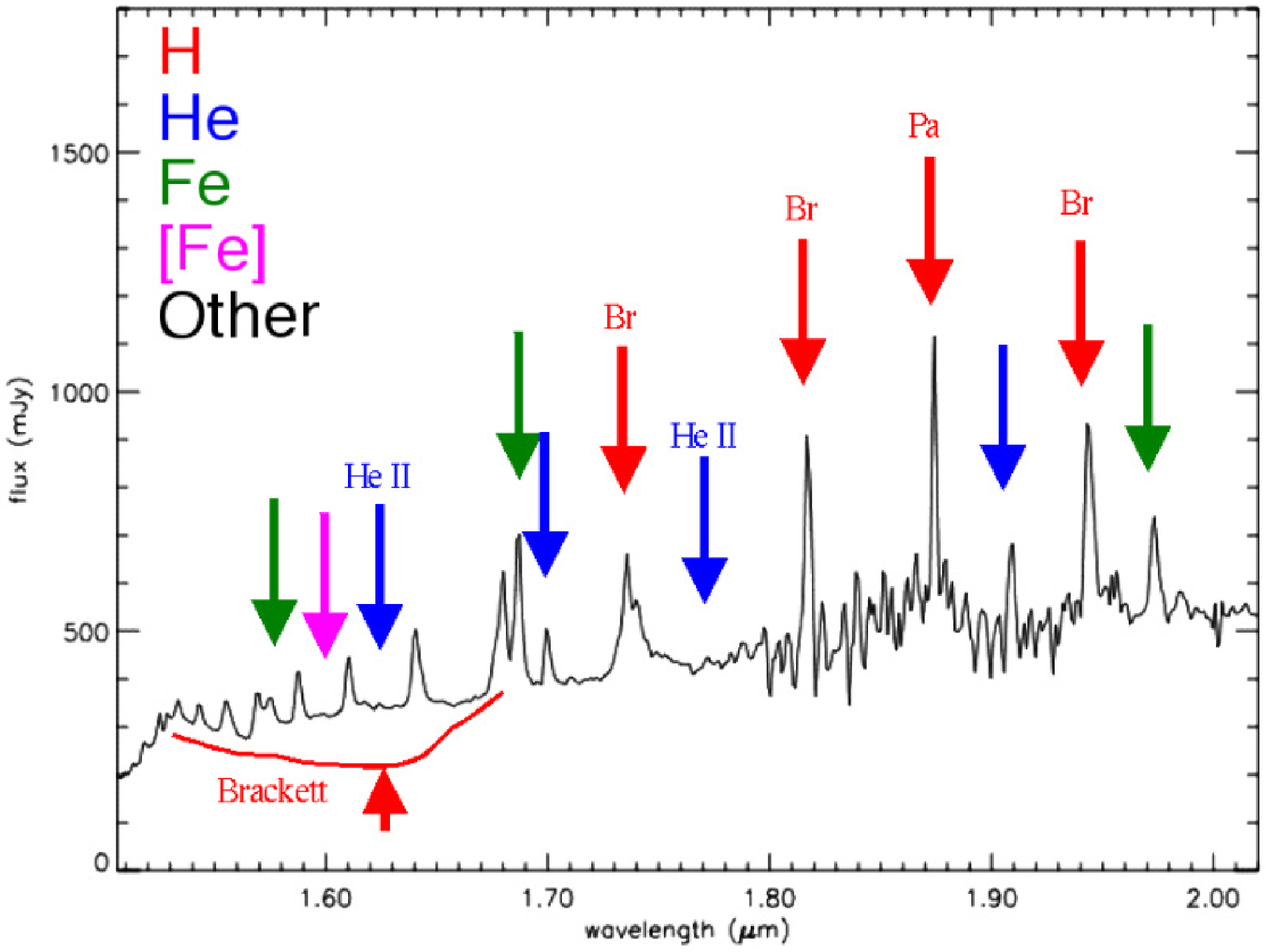}
  \caption{NIR GRF spectrum (1.5-2.05 $\micron$)}
  \label{fig:f16}
\end{figure}
\end{small}

\begin{small}
\begin{figure}
  \includegraphics[width=\columnwidth]{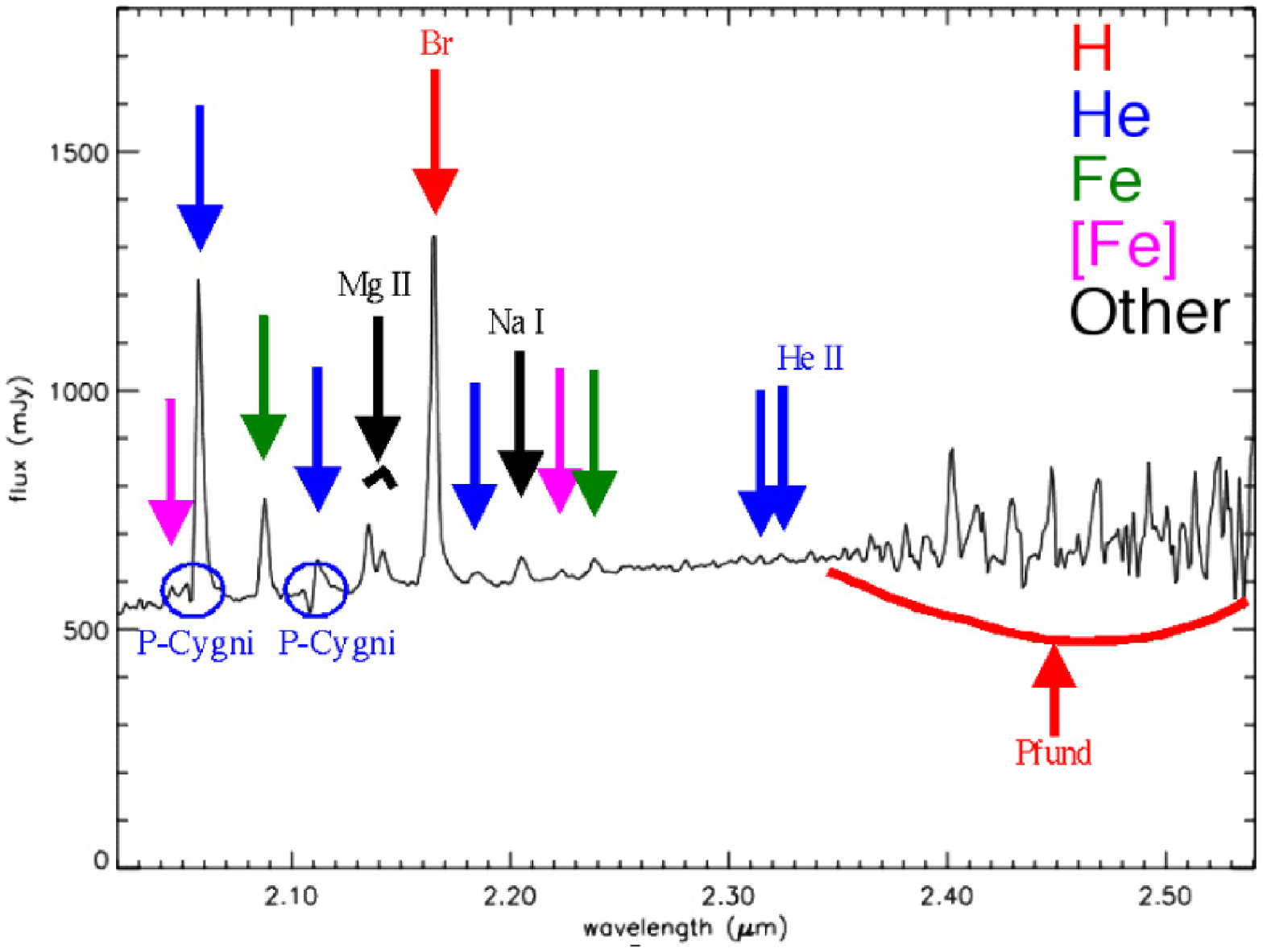}
  \caption{NIR spectrum (2.0-2.55 $\micron$)}
  \label{fig:f17}
\end{figure}
\end{small}

\begin{small}
\begin{figure}
  \includegraphics[width=8.5cm]{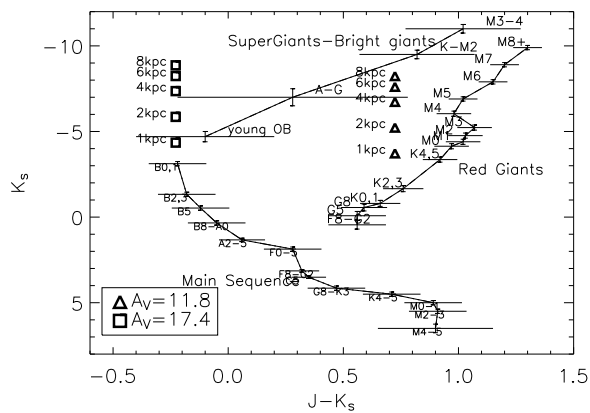}
  \caption{Position on the color-color HR diagram computed from template stars of Ruelas-Mayorga 1991, for distance from 1 to 8 kpc, and for two absorption values: $A_{V}=17.4$ and $A_{V}=11.8$.}
  \label{fig:f21}
\end{figure}
\end{small}

\section*{Acknowledgments}

SC is grateful to Roland Walter and John Tomsick for very helpful
discussions on the nature of the source during this workshop.
SC is also thankful to the ESO panel who understood the utility
of multi-wavelength ToO programmes to reveal the nature of
high-energy sources, and to the ESO staff (especially Malvina
Billeres), who conducted the observations.\\


\end{document}